\documentclass[pre,twocolumn,twoside,showpacs,superscriptaddress,floatfix]{revtex4}
\usepackage{graphicx,amssymb,amsmath,epsfig,color}
\epsfclipon

\begin{document}
 \title{Sample-dependent first-passage time distribution in a disordered medium}
 \author{Liang Luo$^1$ and Lei-Han Tang}
\affiliation{Beijing Computational Science Research Center, Beijing 100094, China}
\affiliation{Department of Physics and Institute of Computational and Theoretical Studies,
Hong Kong Baptist University, Hong Kong}
\date{\today}

\begin{abstract}
Above two dimensions, diffusion of a particle in a medium with quenched random traps is 
believed to be well-described by the annealed continuous time random walk (CTRW).
We propose an approximate expression for the first-passage-time (FPT) distribution in a given sample
that enables detailed comparison of the two problems.
For a system of finite size, the number and spatial arrangement of deep traps yield significant
sample-to-sample variations in the FPT statistics.
Numerical simulations of a quenched trap model with power-law sojourn times 
confirm the existence of two characteristic time scales and a non-self-averaging FPT distribution,
as predicted by our theory.

\end{abstract}

\pacs{87.16.dj, 02.50.-r, 05.40.-a, 87.10.Mn}

\maketitle

\section{Introduction}

The diffusive motion of macromolecules is an essential part of cellular life. It controls
the speed of a large number of cellular processes such as signalling, assembly of protein complexes 
and molecular machines, and exit of mRNAs from the cell nucleus\cite{barkai12,bressloff13,revnucl14}.
Recent advances in {\it in vivo} single-molecule imaging have greatly enriched our
knowledge of thermally driven transport in the heterogeneous intracellular medium.
Experimental measurements have generally indicated a subdiffusive behavior of
fluorescently tagged particles in the cytoplasm and the nucleoplasm, as well as on the plasma membrane\cite{hofling13,golding06,jeon11}. The origin of the observed anomalous phenomenon is still much debated. 

Hitherto, the statistical characteristics of measured particle trajectories are mostly
interpreted using one of the three theoretical models:
the fractal Brownian motion with temporal correlation of particle displacements 
\cite{mandelbrot68,lutz01,weron05,jeon13a,weiss13}, the 
continuous time random walk (CTRW) with power-law sojourn times\cite{montroll65,jeon13}, 
or the obstructed diffusion caused by organelles in the diffusion path\cite{saxton94,havlin87}. 
While these models present different scenarios for the cause of subdiffusive scaling of particle 
displacement with time, they do not take into account the quenched nature of the intracellular medium. 
For diffusive transport across length scales larger or comparable to the size of chromosomes, endoplasmic
reticulum\cite{LiHui15} and other organelles, 
the environment is usually static during the passage of a tracked particle. 
In such a situation, one expects significant
sample-to-sample variations whose statistical mechanical characterization is challenging\cite{metzler14}.

One of the well-known models in this context is the quenched trap model (QTM) 
defined by a set of hopping rates $\tau_i^{-1}$ out of sites $i$ on a $d$-dimensional lattice\cite{machta85,bouchaud90,monthus03,bouchaud03,barkai11prl,luo14}, as illustrated in Fig.~1.
The time constants $\tau_i$ are quenched random variables drawn from a distribution $P_s(\tau)$. 
Above the critical dimension $d_c=2$ for returning walks,
it is generally expected that the QTM has the same scaling properties as the CTRW 
where $\tau_i$ is re-assigned according to the distribution $P_s(\tau)$
upon each visit to site $i$. The argument was formalized in a 
renormalization group analysis by Machta\cite{machta85}.

\begin{figure}
\centering
\includegraphics[width=5.5cm]{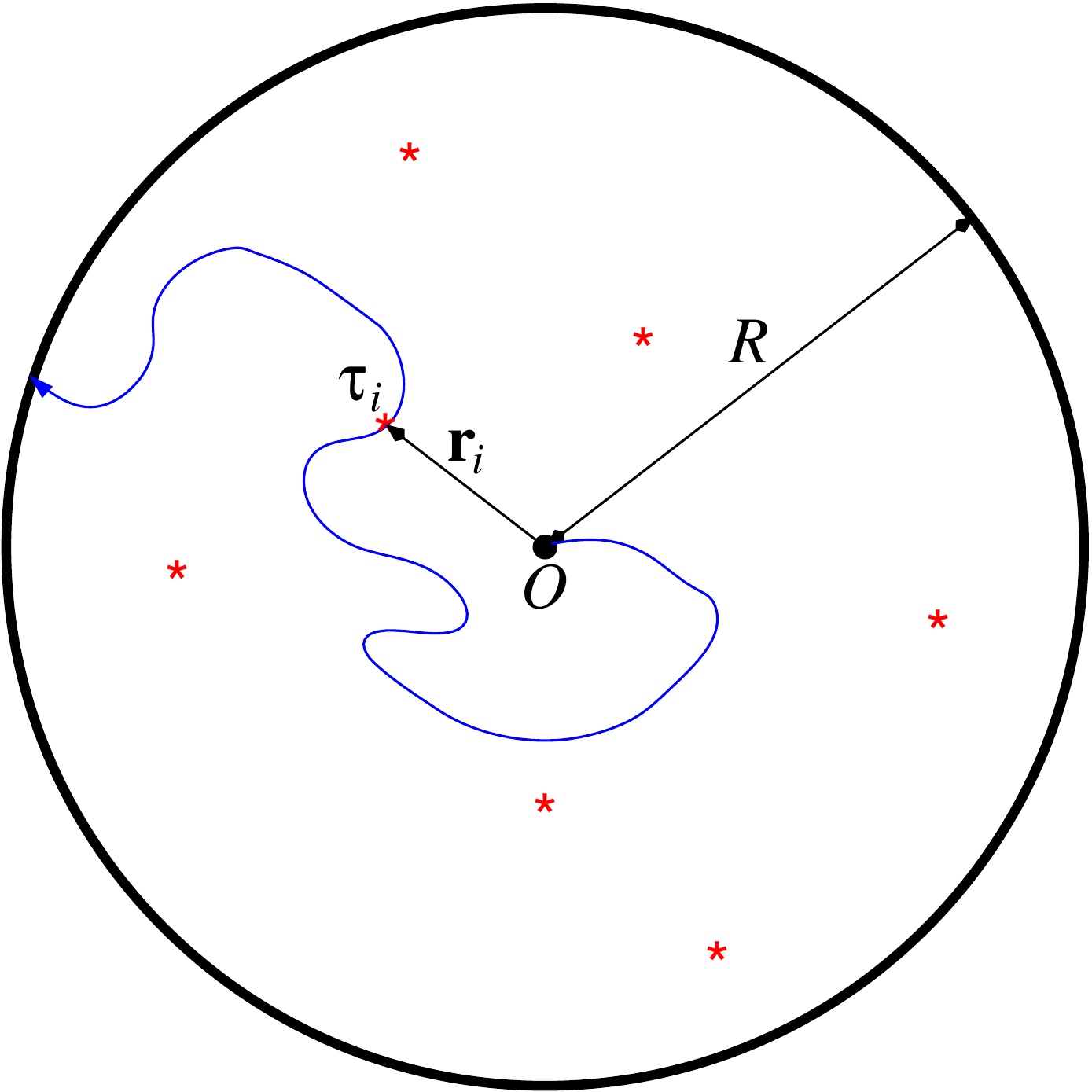}
\caption{\label{trap} (color online). First-passage trajectory (blue line) of a diffusing particle in a quenched environment with deep traps (red dots). }
\end{figure}

Indications that the scaling properties of the QTM and
that of the CTRW may not be identical in high dimensions 
can be found from previous studies of the first-passage-time (FPT) between start and target sites in a confined geometry.
The FPT has been suggested to be an important characteristic 
for understanding reaction kinetics inside a cell\cite{benichou10,krusemann14}.
Consider a simple example illustrated in Fig. 1 where the start site is at the center of a sphere of radius $R$ and
the target is any site on the surface of the sphere.
For this geometry and a power-law distribution $P_s(\tau)\sim \tau^{-\mu-1}$ of trapping times $\tau_i$,
the mean FPT was shown to scale with $R$ as $\tau_Q\sim R^{d/\mu}$\cite{bouchaud90,condamin07pre}.
In comparison, for $\mu<1$, the typical FPT of the corresponding CTRW is given by $\tau_{\rm typ}\sim R^{2/\mu}$,
while the mean FPT diverges. Since the mean FPT tends to be dominated
by rare events when the distribution has a fat tail, the above results suggest qualitatively different 
tail of the FPT distribution in the two models.

In this paper, we develop an analytic scheme to characterize
the full FPT statistics in the QTM, and to investigate its sample-to-sample fluctuations.
Simple physical arguments are presented to reveal the subtle differences in the
rare-event statistics between the quenched and annealed systems. 
The two time scales, $\tau_{\rm typ}\sim R^{2/\mu}$ for the longest trapping time
on a typical FPT trajectory, and $\tau_Q\sim R^{d/\mu}$ for the longest trapping time in the entire system,
arise naturally from the discussion. Based on these understandings, we propose
a decomposition scheme that expresses the FPT distribution as an ordered
sum of exponentials associated with deepest traps in a given sample.
Variations in the strength and location of these traps give rise to sample-to-sample fluctuations 
in the tail of the FPT distribution which we characterize analytically.
The analytic results are shown to be in excellent agreement with
numerical simulations of the two models. 

The paper is organized as follows. In Sec. II we define the Quenched Trap Model and
mention a few of its basic properties. Section III presents our analytic calculations of the FPT distribution
from the point of view of rare-event statistics. Section IV contains results from numerical simulations
of the two models, highlighting sample-to-sample variations in the QTM. A brief summary is given
in Sec. V.

\section{The Quenched Trap Model}

The Quenched Trap Model can be formulated as a Markov process for a diffusing particle
whose states are sites on a $d$-dimensional hypercubic lattice with a lattice constant $a$. 
The transition rate from site $i$ to a neighboring site $j$ is given by $W_{i\rightarrow j}=(2d)^{-1}\tau_i^{-1}$. 
The hopping rate $k_i=\tau_i^{-1}$ out of site $i$
can be associated with a site energy $V_i (<0)$ through the Arrhenius law $k_i=\omega_0\exp(V_i/T)$, where 
$\omega_0$ is the attempt rate and $T$ the ambient temperature. Of special interest is when the site energies
$V_i$ is exponentially distributed,
\begin{equation}
\label{gumbel}
P(V)=T_g^{-1}e^{V/T_g}.\qquad (V<0)
\end{equation}
Consequently, the distribution of $\tau_i$ follows a power-law,
\begin{equation}
\label{sft}
P_s(\tau)=\mu\omega_0^{-\mu}\tau^{-\mu-1}.\qquad (\tau>\omega_0^{-1})
\end{equation}
Here the exponent $\mu=T/T_g$. Below we will choose $\tau_0\equiv\omega_0^{-1}$ as the unit of time 
unless otherwise specified.

The QTM with a power-law distribution of sojourn time constants (\ref{sft}) offers a plausible description of
the diffusion of a macromolecule in the aqueous cellular environment. 
On short time scales, the molecule is trapped in a certain volume of its size due to either
nonspecific binding or cage effect as in colloidal glass. It has been argued that the ``Gumbel distribution''
approximated by (\ref{gumbel}) is often encountered when multiple factors of
comparable strength contribute to the trapping energy $V_i$\cite{tang01}. 
On longer time scales, after breaking off from the trap, the particle diffuses normally until it falls into another trap. 
In this scenario, the path taken by the molecule is simply a normal random walk, although the journey time
may have a very broad distribution.
The lattice constant $a$ can be taken to be the linear size of a correlated volume
in the medium, beyond which the local potential $V_i$ on the diffusing particle changes to a substantially
different value. The time constant $\tau_0$ should be chosen accordingly so that (\ref{sft}) provides an adequate
description for hopping between neighboring sites.

According to Eq. (\ref{gumbel}), the site energies are not bounded from below. 
However, in a given realization of the disorder, there will be a deepest trap at energy $V_{\rm Q}=\min_i\{V_i\}$.
Let ${\cal V}\simeq R^d$ be the volume of the system and ${\cal N}={\cal V}/a^d$ be the number of
independent sites. The cumulative distribution of $V_{\rm Q}$ is given by
\begin{eqnarray}
C_V(V_{\rm Q},{\cal N})&=&1-(1-e^{V_{\rm Q}/T_g})^{\cal N}\simeq 1-\exp(-{\cal N}e^{V_{\rm Q}/T_g})\nonumber\\
&=&\hat{C}_V\bigl({V_{\rm Q}\over T_g}+\ln {\cal N}\bigr),
\label{C_VN}
\end{eqnarray}
where the scaling function $\hat{C}(u)=1-\exp(-e^u)$.
Therefore the mean of $V_{\rm Q}$ decreases logarithmically with ${\cal N}$ but its variance remains constant.
Consequently, the energy difference between the deepest and the second deepest
trap, which can be approximated by the difference in $V_{\rm Q}$ in
two independent realizations of the disorder, is finite and independent of ${\cal N}$.
The corresponding trapping time constant 
\begin{equation}
\tau_{\rm Q}(R)=\tau_0\exp(-V_Q/T)\simeq\tau_0(R/a)^{d/\mu}
\label{tau_Q}
\end{equation}
is greater than the second longest time by a factor. 

\section{Analytic calculation of the first-passage-time distribution}

For diffusion-limited biochemical reactions inside a cell, one is interested
in the FPT of a molecule from its birth place to the reaction site. 
In this work, we shall focus on the statistics of the FPT for individual cells.
To simplify the discussion, we shall adopt the simplest geometry as illustrated by Fig. 1,
where the molecule is launched from the origin at $t=0$,
and examine the distribution $F(t,R)$ of the FPT $t$ to an enclosing spherical surface of radius $R$.
Nevertheless, our approach can be adapted to more general geometries 
as considered by B\'{e}nichou {\it et al.} \cite{benichou10} using results from the lattice 
random walk\cite{montroll56,redner07}.

\subsection{The Continuous Time Random Walk}

If the sojourn times of the diffusing particle at a given site are not distributed in a site-dependent manner but
instead follow a common distribution such as Eq. (\ref{sft}), the corresponding stochastic process is known as
the Continuous Time Random Walk. The CTRW can be considered as an annealed version of the QTM, with the
important difference that the statistics of the journey time is independent of the path taken. 
This  ``subordination" property allows one to write,
\begin{equation}
F_{\rm CTRW}(t,R)=\sum_N A_N(R) f_N(t),
\label{subordination}
\end{equation}
where $A_N(R)$ is the probability for a lattice random walker to reach the boundary for the first time in $N$ steps,
and $f_N(t)$ is the probability distribution function of the journey time $t_N=\sum_{i=1}^N t_i$.
Under the continuum approximation at large $N$ (which plays the role of time) and $R$, 
the lattice random walk is described by the diffusion
equation which, under the substitution $x\rightarrow Rx$ and $N\rightarrow R^2 N$, yields 
the scaling solution,
\begin{equation}
A_N(R)=R^{-2}\hat{A}_d(NR^{-2}),
\label{A_scaling}
\end{equation}
where the FPT probability density function (PDF) $\hat{A}_d(u)$ is peaked around $u_{\rm max}\simeq 1$ and decays
exponentially at large $u$. On the other hand, $f_N(t)$ exhibits
a fat tail when the sojourn times $t_i$ are distributed according to (\ref{sft}).

A particularly interesting case is $\mu<1$, where the mean sojourn time
$t_s=\int_0^\infty tP_s(t) dt$ diverges. In such a situation, the sum $t_N$ is dominated by the largest term 
$t_{\rm max}$ whose typical value grows faster than linear in $N$. The latter can be seen from the 
cumulative distribution of $t_{\rm max}$,
\begin{eqnarray}
\label{prob_t_max}
{\rm Prob}(t_{\rm max}>t)&=&1-\prod_{i=1}^N{\rm Prob}(t_i<t)=1-(1-t^{-\mu})^N\nonumber\\
&\simeq& 1-\exp(-Nt^{-\mu}).
\end{eqnarray}
In Appendix A we show that, in this case, the tail of
the distribution of $t_N$ becomes identical to that of $t_{\rm max}$,
\begin{equation}
\label{tauextre}
f_N(t)\approx P_N(t_{\rm max}=t)=N\mu(1-t^{-\mu})^{N-1}t^{-\mu-1}.
\end{equation}

Equations (\ref{prob_t_max}) and (\ref{tauextre}) show that the distributions of $\tau_{\rm typ}$ and of $t_N$ are both
peaked around $\tau_{\rm typ}\simeq N^{1/\mu}$ and have the expected power-law tail beyond $\tau_{\rm typ}$.
Combining Eqs. (\ref{subordination})-(\ref{tauextre}), we obtain the subdiffusive scaling\cite{bouchaud90,condamin07prl},
\begin{eqnarray}
F_{\rm CTRW}(t,R)&\simeq&\int {dN\over R^2}\hat{A}_d\Bigl({N\over R^2}\Bigr)N\mu\exp\Bigl(-{N\over t^\mu}\Bigr)t^{-\mu-1}\nonumber\\
&=&\mu R^2t^{-\mu-1}\Phi(tR^{-2/\mu}).\qquad (\mu<1)
\label{F-scaling}
\end{eqnarray}
Here $\Phi(z)=\int_0^\infty du u \hat{A}_d(u)\exp(-u/z^\mu)$ increases
monotonically with $z$ and saturates to $\Phi_\infty$ at large $z$.
Hence $F_{\rm CTRW}(t,R)$ is peaked at 
\begin{equation}
\tau_{\rm typ}(R)\simeq \tau_0(R/a)^{2/\mu}.
\label{t_typ}
\end{equation}
The corresponding cumulative FPT distribution takes the form,
\begin{equation}
\label{canneal}
C_{\rm CTRW}(t,R)=\int_t^\infty dt' F(t',R)= \hat{C}(tR^{-2\mu}).
\end{equation}
Here $\hat{C}(z)=\int_z^\infty dy \mu y^{-\mu-1}\Phi(y)$. For $z\gg 1$, $\hat{C}(z)\simeq \Phi_\infty z^{-\mu}$.

\subsection{The FPT distribution in the QTM}

Equation (\ref{subordination}) integrated over $t$ is a special example of a general formula for the 
cumulative FPT distribution,
\begin{equation}
C(t,R)=\sum_\Gamma W_\Gamma C_\Gamma(t),
\label{decomposition}
\end{equation}
where the summation extends over all possible lattice walks $\Gamma$ connecting the launch site to the boundary.
Here $W_\Gamma$ is the probability for a path $\Gamma$ in the lattice walk, 
with the normalization $\sum_\Gamma W_\Gamma=1$,
and $C_\Gamma(t)$ the probability that the total passage time exceeds $t$. 
In terms of the sojourn time distribution $P_i(t)$ on site $i$, we may write
\begin{equation}
C_\Gamma(t)=\int H\bigl(t-\sum_{i\in\Gamma}t_i\bigr)\prod_{i\in\Gamma} P_i(t_i)dt_i,
\label{C-prod}
\end{equation}
where $H(t)$ is the Heaviside step function.

For the CTRW, the sojourn time distributions $P_i(t)$ are identical for all sites.
Hence $C_\Gamma(t)$ depends only on the path length $N$, in which case
paths with the same $N$ can be grouped together, leading to Eq. (\ref{subordination}).
On the other hand, $C_\Gamma(t)$ for the QTM depends on the actual sites visited. 
An interesting question is whether the disordered averaged $C_\Gamma(t)$ 
depends only on the path length $N_\Gamma$ but not its spatial trajectory. 
If so, the ensemble-averaged $C(t,R)$ can again be cast in the form of Eq. (\ref{subordination}).

The answer to the above question is in the affirmative if, on the lattice walk $\Gamma$, 
each site visited appears only once, i.e., there is no return to any given site on the path.
For this group of lattice walks, the disorder-averaged $C_\Gamma(t)$ is equivalent to
its annealed counterpart whose sojourn time distribution is given by,
\begin{equation}
\langle P_i(t)\rangle=\int_1^\infty {e^{-t/\tau}\over\tau}{\mu d\tau\over\tau^{\mu+1}}
=\mu t^{-\mu-1}\int_0^t x^\mu e^{-x}dx.
\label{P_anneal}
\end{equation}
Here and elsewhere we use $\langle\cdot\rangle$ to denote average over the quenched disorder.
It is easy to verify that Eq. (\ref{P_anneal}) also
has a fat tail that decays as $t^{-\mu-1}$ at large $t$. 

The theory of lattice random walks\cite{montroll56,redner07} can be applied to calculate the distribution of
returns for the setup illustrated in Fig. 1. For $d<2$, the typical number of returns grows with $R$ as $R^{2-d}$
(logarithmic at $d=2$). Hence a different scheme to compute the ensemble average of $C(t,R)$ is required.
On the other hand, for $d>2$, the return distribution decays exponentially.
Furthermore, the majority of returns to a given site take place in a short section of the walk, allowing for
a renormalization group treatment of their effects on $C_\Gamma(t)$\cite{machta85}. 
In the following, however, we will take a different route to reorganize terms in Eq. (\ref{decomposition}) to
compute $C(t,R)$, focusing on contributions from the deepest traps.

The calculation in Appendix A shows that $C_\Gamma(t)$ has an exponential tail with a time constant
$\tau_\Gamma=\sum_{i\in\Gamma}\tau_i$. Furthermore, under Eq. (\ref{sft}) at $\mu<1$, 
$\tau_\Gamma$ is dominated by the largest trapping constant $\tau_{\rm max,\Gamma}={\rm max}_{i\in\Gamma}\{\tau_i\}$.
These observations suggest that $C_\Gamma(t)$ 
may be replaced by the cumulative sojourn time distribution at the deepest trap on path $\Gamma$.
More precisely, labelling each path $\Gamma$ by $\tau_{\rm max,\Gamma}$,
we rewrite Eq. (\ref{decomposition}) as,
\begin{equation}
C(t,R)=\sum_i C_i(t,R),
\label{C_i}
\end{equation}
where the partial sum
\begin{equation}
C_i(t,R)=\sum_{\Gamma, i\in\Gamma,\tau_i=\tau_{\rm max,\Gamma}}W_\Gamma C_\Gamma(t)
\label{C_i_Gamma}
\end{equation}
is restricted to paths $\Gamma$ that go through site $i$ with $\tau_i$ the largest trapping constant on $\Gamma$.

To proceed further, we introduce a single-trap model where $\tau_j=\tau_0$ for all sites except at site
$i$ located a distance $r$ from the center of the sphere, for which $\tau_i=\tau_{\rm Q}\gg \tau_0(R/a)^2$.
In Appendix B, we present a calculation of the probability that the trap is visited by a 
first passage trajectory illustrated in Fig. 1,
\begin{equation}
\label{w0asymp-eq}
w(r,R)\simeq(1-f_d){2d\over S_d}{r^{2-d}-R^{2-d}\over d-2}. \quad (1\ll r< R)
\end{equation}
Here $f_d$ is the probability of return on an infinite lattice and $S_d=2\pi^{d/2}/\Gamma(d/2)$ is the surface area
of unit sphere. For simple cubic lattice, $f_3\simeq 0.34$. Taking into account multiple visits\cite{montroll56}, the
calculation yields an exponential tail for the cumulative FPT distribution
with a renormalized time constant, 
\begin{equation}
\label{cqexp}
C_1(t,R\vert\tau_{\rm Q},r)\simeq w(r,R)\exp(-e_d t/\tau_{\rm Q}),
\end{equation}
where $e_d\equiv 1-f_d$ is the probability of no return. 
Contributions from paths that do not go through the trap die out at times much shorter than $\tau_{\rm Q}$.

We now compare Eq. (\ref{C_i_Gamma}) with (\ref{cqexp}) when the site $i$ corresponds to the deepest trap
in the system, with $\tau_i=\tau_Q$.
In this case, the sum in Eq. (\ref{C_i_Gamma}) includes all paths that go through site $i$, as in the single-trap model.
The total time spent on other sites on the path is of the order of $\tau_{\rm typ}(R)$ in Eq. (\ref{C_i_Gamma}) and
$\tau_0(R/a)^2$ in the single-trap model. Hence, in both cases, the total passage time is well approximated by
the sojourn time on the deepest trap. This allows us to write, for the deepest trap,
\begin{equation}
C_i(t,R)\simeq C_1(t,R|\tau_i,r_i),
\label{C_iC_1}
\end{equation}
where $r_i$ is the distance of site $i$ to the origin.

For the second deepest trap in the system, the same argument leading to (\ref{C_iC_1})
applies except that the right-hand-side of the equation 
contains extra contributions from paths that visit both the deepest and the second deepest trap. 
To eliminate such terms, we need to replace $w(r,R)$ in Eq. (\ref{cqexp}) by $w(i,j,R)$
that gives the probability of reaching the boundary via site $i$ with an absorbing site at $j$.
Continuing the procedure to the $(n+1)$th deepest trap, 
one needs to compute the visit probability to the site in question in the presence of $n$ absorbing sites
and the system boundary, which is quite challenging. 
However, using the annealed approximation that, in each step,
the walker has a probability $p=n/{\cal N}$ to run into one of these sites,
we obtain the survival probability $(1-p)^N\simeq \exp(-nN/{\cal N})$ in an $N$-step walk. 
Applying this estimate to first passage trajectories with $N\simeq R^2$, we see that
Eq. (\ref{C_iC_1}) holds approximately when $n<n_c={\cal N}/N\simeq (R/a)^{d-2}$,
but for weaker traps, $C_i(t,R)$ diminishes rapidly. The trapping time constant
at $n_c$ satisfies $(\tau/\tau_0)^{-\mu}=n_c/{\cal N}=(R/a)^2$, yielding a cut-off time constant
$\tau_{\rm typ}(R)$ given by Eq. (\ref{t_typ}).

The above discussion yields the following approximate expression for the probability that the
FPT in a given sample is greater that $t$,
\begin{equation}
C_{\rm QTM}(t,R)\simeq \sum_{i,\tau_i>\tau_{\rm typ}(R)}C_1(t,R\vert\tau_i,r_i),
\label{C_superposition}
\end{equation}
where $C_1(t,R\vert\tau_i,r_i)$ at $t>\tau_{\rm typ}(R)$ is given by Eq. (\ref{cqexp}). 
Each term in the sum decays exponentially with a time constant given by the strength of the trap.
The largest time constant is set by the deepest trap in the system.

Given the simple mathematical form of Eq. (\ref{C_superposition}), various properties of the QTM can
be derived analytically. For example, as we show in Appendix C,
the mean and variance of $C(t,R)$ over different disorder realizations are easily computed.
For $2<d<4$, the results are given by,
\begin{eqnarray}
\langle C(t,R)\rangle&\simeq&\mu e_d^{1-\mu}R^2t^{-\mu}\gamma\Bigl(\mu,{e_dt\over R^{2/\mu}}\Bigr),
\label{C_mean}\\
\langle[\Delta C(t,R)]^2\rangle&\simeq&
\mu (2e_d)^{2-\mu}S_d^{-1}{d\over 4-d}R^{4-d} t^{-\mu}\nonumber\\
&&\quad\times\gamma\Bigl(\mu,{2e_d t\over R^{2/\mu}}\Bigr).\label{C_var}
\end{eqnarray}
Here $\gamma(\mu,z)=\int_0^zdx x^{\mu-1}e^{-x}$ is the incomplete gamma function.
As expected, the ensemble averaged FPT distribution (\ref{C_mean}) 
is essentially the same as that of the CTRW. 
On the other hand, its relative fluctuation satisfies the scaling,
\begin{equation}
\label{fss}
{\langle[\Delta C(t,R)]^2\rangle^{1/2}\over \langle C(t,R)\rangle}=R^{-(d-2)/2}f(tR^{-2/\mu}),
\end{equation}
where $f(z)\sim z^{\mu/2}$ for $z\gg 1$. The normalized fluctuation is proportional to
$(t/\tau_{\rm Q})^{\mu/2}$ in the intermediate time regime $\tau_{\rm typ}<t<\tau_{\rm Q}$, 
and becomes of order 1 or bigger when $t$ exceeds $\tau_{\rm Q}$.

\section{Simulation results}

We performed extensive kinetic Monte Carlo simulations of the CTRW and the QTM on the three-dimensional 
simple cubic lattice to verify the analytic results presented in Sec. III.
In each run, a particle is released from the origin at $t=0$
and performs unbiased random walk through nearest neighbor hops. 
The walk is terminated when the particle, for the first time, reaches a site at a distance greater than $R$
from the origin. The total passage time is given by the sum of sojourn times at each stop along the path.
In the case of the CTRW, the sojourn times are drawn independently from the distribution (\ref{sft}).
For the QTM, a set of trapping time constants $\tau_i$ are first assigned to the lattice sites. The actual
sojourn time upon each visit to a given site follows an exponential distribution with the pre-assigned time constant.
The system sizes investigated are from $R=7$ to 15. 

\subsection{The FPT distribution}

Figure 2 shows three examples for the cumulative distribution function $C(t,R)$
at $R=7$ and $\mu=0.71$. Here $\tau_{\rm typ}=R^{2/\mu}=240$ in units chosen.
More than $10^6$ trajectories are generated to obtain accurate statistics.
As seen from the figure, for both CTRW and the QTM, $C(t,R)$ begins to drop from its maximum value 1
around $\tau_{\rm typ}$. The shape of $C(t,R)$ from the two samples in the QTM is quite similar for $t$
around $\tau_{\rm typ}$, confirming that the most probable value of the FPT in the QTM does not vary
significantly from sample to sample and coincides with its annealed counterpart CTRW.
In the tail part of the FPT distributions, however, the two samples in the QTM show progressively
larger deviations from the power-law $C(t,R)\simeq R^2t^{-\mu}$ (dash-dotted line)
that describes well the CTRW data.  
At very long times, $C(t,R)$ from the QTM exhibits the exponential decay predicted by the theory
presented in Sec. III. The tail part of the FPT distribution in each case is well-described by Eq. (\ref{C_superposition}).

\begin{figure}
\centering
\includegraphics[width=8cm]{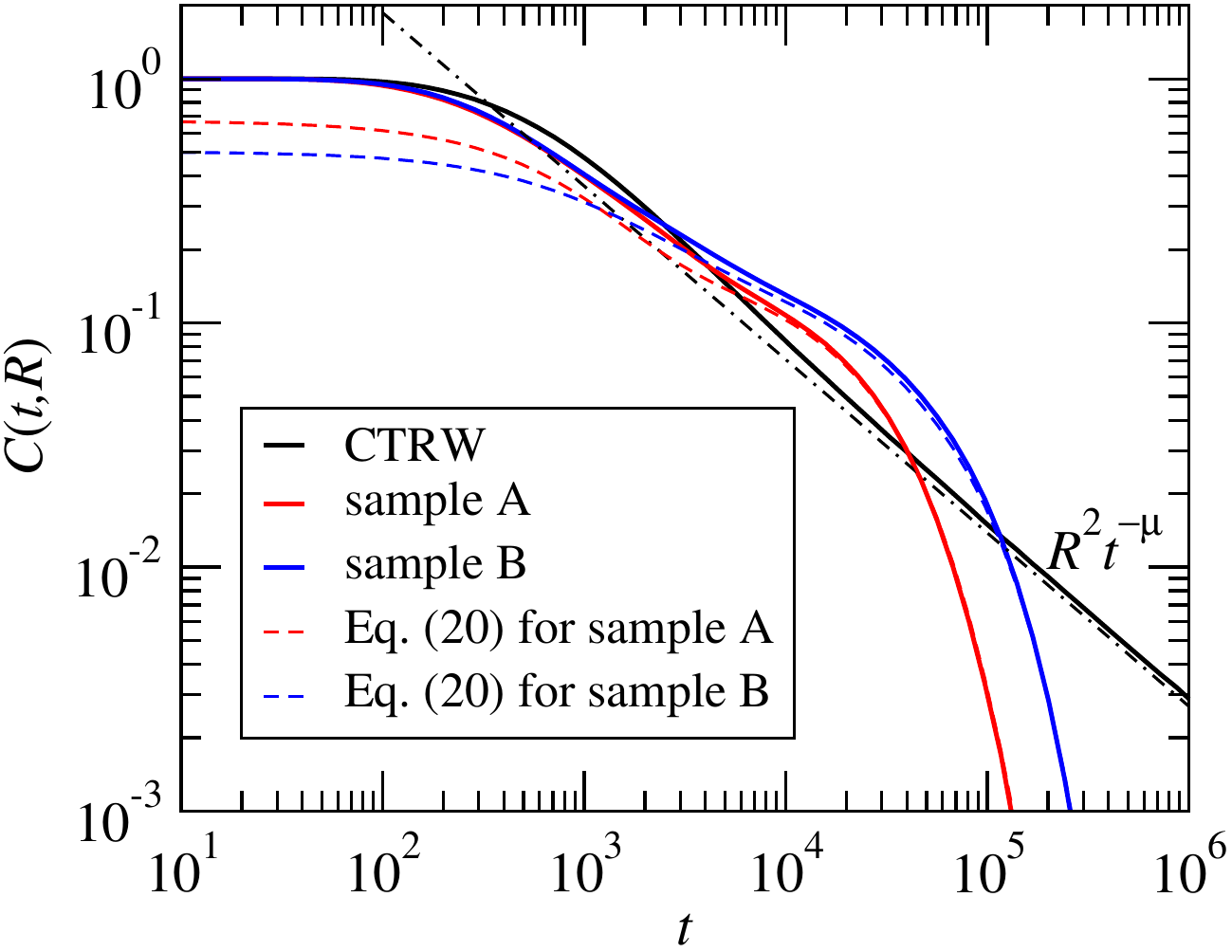}
\caption{\label{qdist}(color online). Cumulative distributions of the first passage time from simulations of the CTRW model 
and the QTM on the three-dimensional simple cubic lattice. 
Here $R=7$ and $\mu=0.71$. Dashed lines are
computed using Eq.~(\ref{C_superposition}) for the specific trap configuration
in each of the two samples, respectively. }
\end{figure}

Figure 3 shows our simulation data for the disorder-averaged $C(t,R)$ and its relative fluctuation from 
10,000 realizations of the QTM at the three system sizes $R=7$, 11 and 15. 
Also shown are numerical evaluations of the corresponding analytic expressions presented in Sec. III.
Excellent data collapse upon scaling of the variables is seen over six decades in time.
Both Eqs. (\ref{C_mean}) and (\ref{fss}) are in quantitative agreement with simulation data on the tail side
of the FPT distribution. 
For $t<\tau_{\rm typ}\simeq R^{2/\mu}$, Eq. (\ref{C_mean}) yields a value $e_d\simeq 0.66$ less than 1,
presumably due to contributions from trajectories not included in the sum (\ref{C_superposition}).
The latter is also responsible for the discrepancy between Eq. (\ref{fss}) and the simulation data in the short
time regime, where the actual sample-to-sample variations of $C(t,R)$ decrease rapidly due to self-averaging.
Taken together, the simulation results confirm quantitatively the decomposition scheme
Eq. (\ref{C_superposition}) for the tail of the FPT distribution in the QTM.

\begin{figure}
\includegraphics[width=8cm]{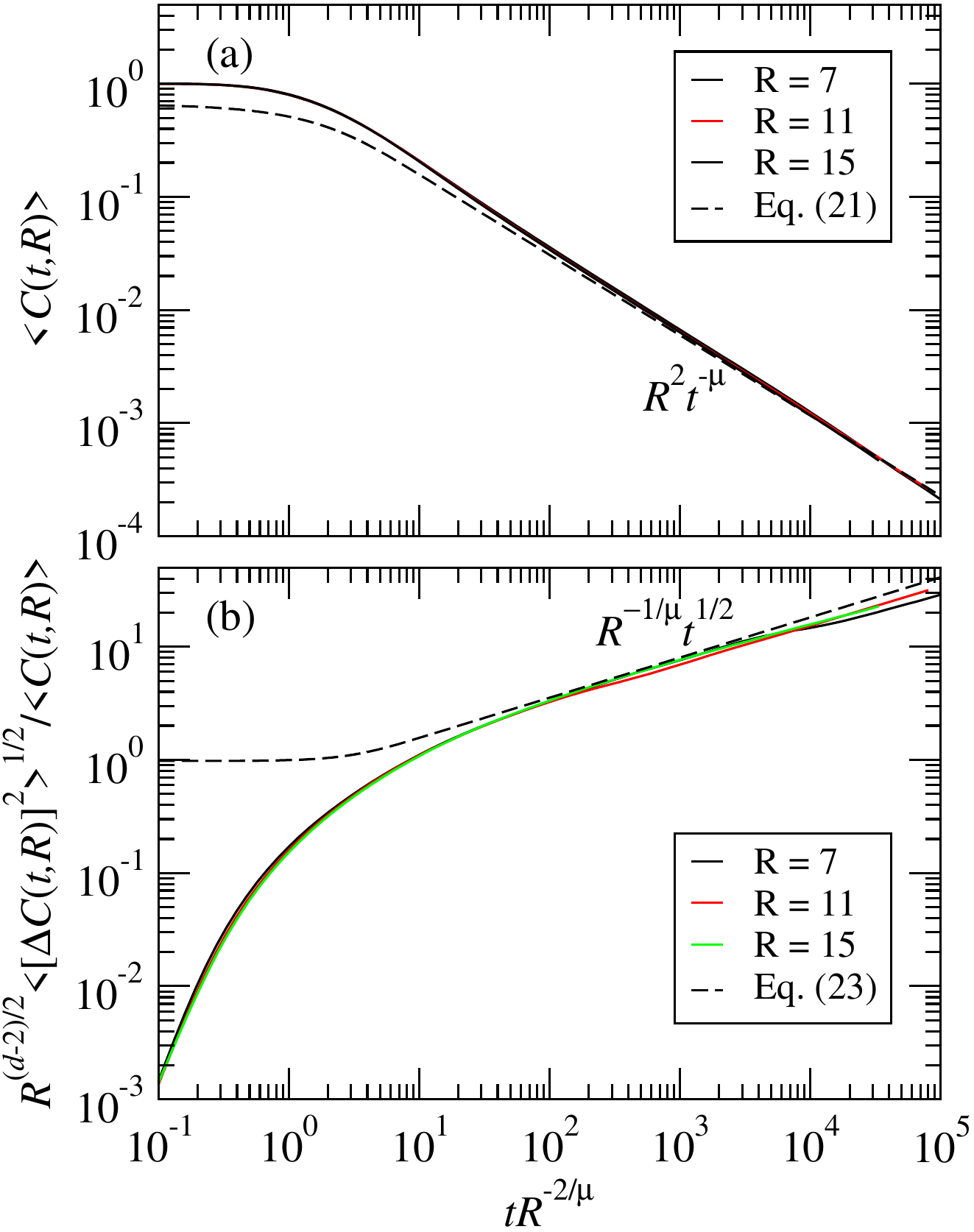}
\caption{\label{c-statistics}(color online). 
Simulation data of the QTM on the three-dimensional simple cubic lattice at $\mu=0.71$.
(a) Scaling plot of the mean cumulative FPT distribution for
three different system sizes $R=7$ (black), 11 (red) and 15 (green), together with
numerical evaluation of Eq. (\ref{C_mean}) (dashed line).
(b) Scaling plot of the relative fluctuation of the cumulative FPT distribution 
from simulations of the QTM at three different sizes, along with numerical evaluation of
Eq. (\ref{fss}).
}
\end{figure}

\subsection{Hit map}

We have also monitored the statistics of the end position of the first passage trajectories as
a function of the passage time $t$. This defines a time-dependent hit map on the absorbing surface.
For the CTRW, since the trajectories are spatially independent of each other,
the hit map is uniform at all times apart from statistical fluctuations
when only a finite number of hits are registered. However, the QTM is expected to show
a correlated hit density pattern which evolves over time. Sites
with long sojourn time constants cause delay to trajectories going through them, 
thereby casting their shadows on the absorbing surface at short times. At very long times, however,
all lattice random walks will have sufficient time to complete their journey, and uniformity is restored.

To construct the spatially-resolved landing probability on the absorbing surface, we first record
$H({\bf x}\vert t,R)$, the number of hits registered at site ${\bf x}$ up to time $t$.
Since a regular lattice is used in the simulations, different surface sites have a different
cross section to capture the incoming walkers.
To minimize this effect, we performed local averaging by collecting hits into a neighborhood
instead of a single site on the boundary. The neighborhood $\Gamma_{\bf x}$ of a given boundary site {\bf x} 
includes the site itself plus 
its nearest-neighbors and next nearest-neighbors who are also boundary sites. 
Let $N_{\bf x}$ be the number of boundary sites in $\Gamma_{\bf x}$.
A coarse-grained hit number is defined as,
\begin{equation}
H_\text{CG}({\bf x}\vert t,R)={1\over N_{\bf x}}\sum_{{\bf x}'\in\Gamma_{\bf x}}H({\bf x}'\vert t,R),
\label{coarse-graining}
\end{equation}
Using the coarse-grained data, we further normalize against $H_{\rm CG}({\bf x}|t=\infty,R)$ 
to remove the lattice effect, i.e., by introducing the normalized cumulative hits,
\begin{equation}
h({\bf x}|t,R)\equiv {H_{\rm CG}({\bf x}\vert t,R)\over H_{\rm CG}({\bf x}\vert \infty,R)}.
\label{normalized-hit}
\end{equation}

\begin{figure}
\includegraphics[width=8cm]{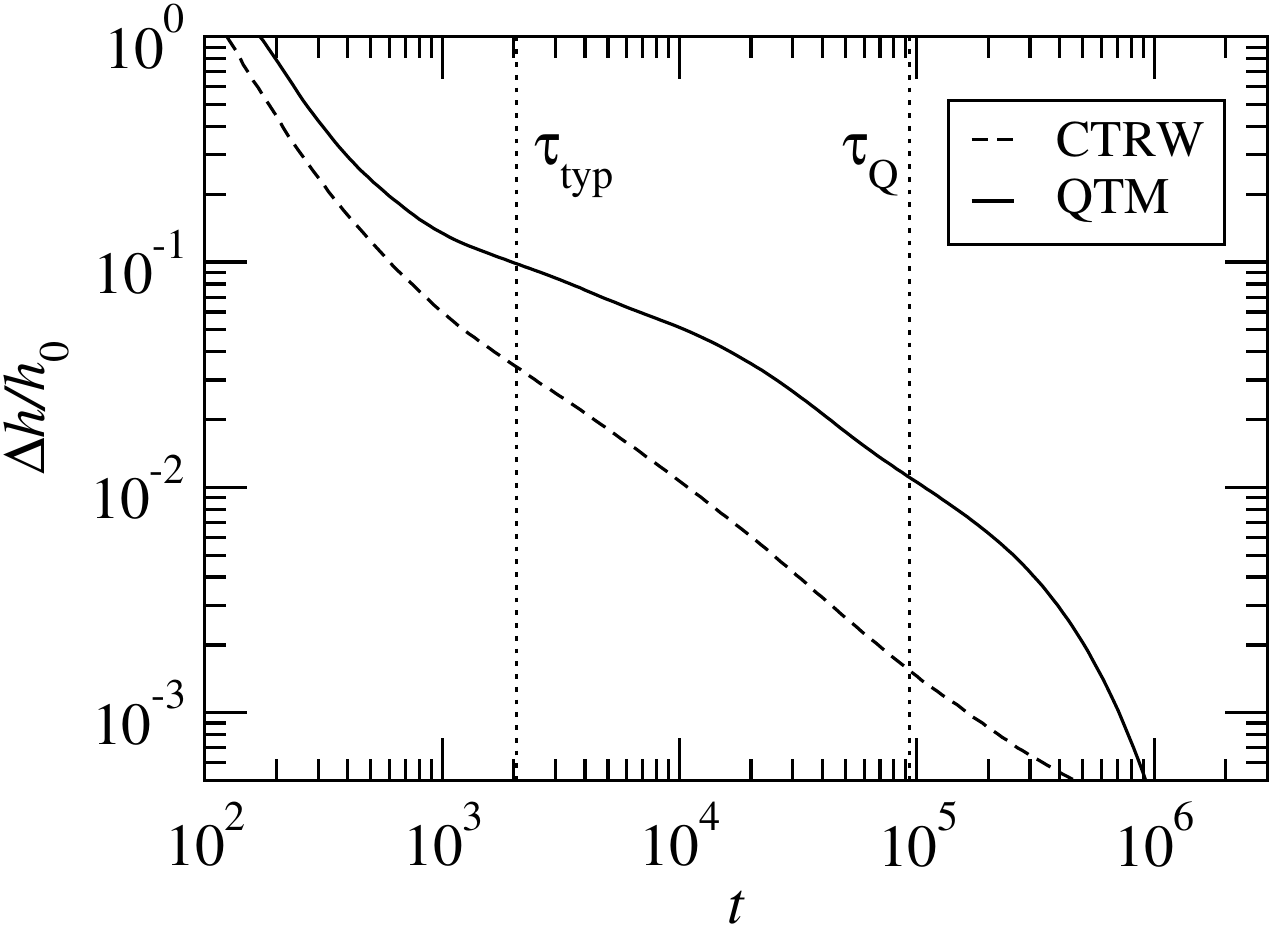}
\caption{\label{spatialfluc} Relative hit density fluctuation in the CTRW and QTM. Here $R=15$
and $\mu=0.71$. }
\end{figure}

Figure \ref{spatialfluc} shows the relative spatial fluctuation of $h({\bf x}|t,R)$ in a system of
size $R=15$ by taking statistics over $10^7$ launches.  Here 
\begin{eqnarray}
h_0(t,R)&=&\langle h({\bf x}\vert t,R)\rangle_{\bf x},\\
\Delta h(t,R)&=&[\langle h^2({\bf x}\vert t,R)\rangle_{\bf x}-h_0^2(t,R)]^{1/2},
\label{delta-h-fluc}
\end{eqnarray}
where $\langle\cdot\rangle_{\bf x}$ denotes average over all boundary sites.
For both the QTM and the CTRW, the relative fluctuation decreases as the number of hits accumulate.
It is also evident that the spatial inhomogeneity is much stronger in the QTM than in the CTRW.
Ideally, one expects the hit pattern to saturate around $\tau_Q$ when most of the walkers have completed
their journey to the absorbing surface. Equation (\ref{normalized-hit}) somewhat under-estimates this 
terminal fluctuation by adopting $H_{\rm CG}({\bf x}\vert\infty, R)$ as the normalization.
The latter contains both the lattice effect as well as residual fluctuations from either the
quenched disorder or a finite number of launches.

\begin{figure}
\centering
\includegraphics[width=8.8cm]{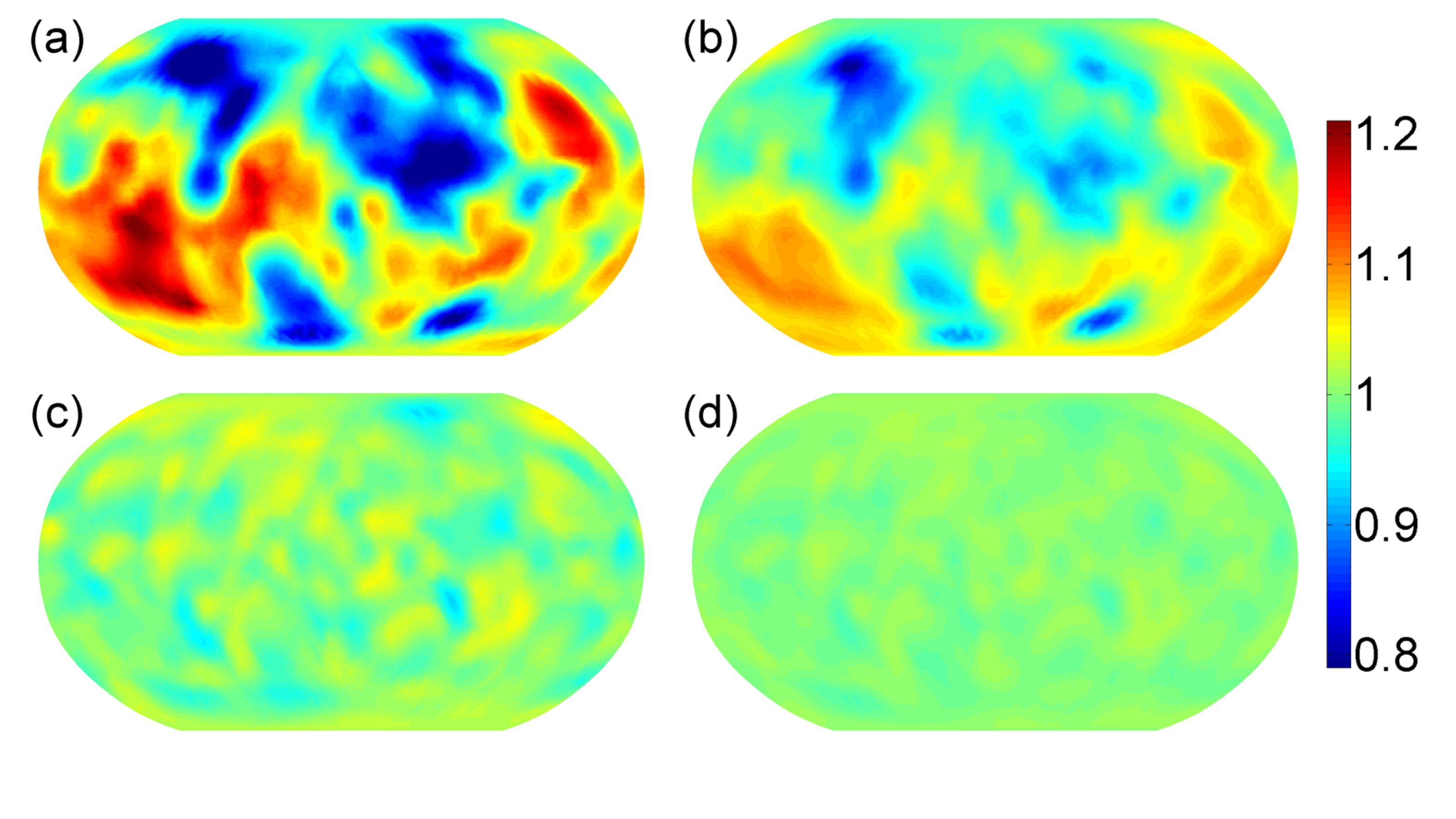}
\caption{\label{worldmap} (color online). Earth plot of the normalized hit map on the absorbing surface at $R=15$.
Here $\mu=0.71$. (a) one sample of the QTM at $t=6200$. (b) same sample at $t=20 000$.
(c) CTRW at $t=6200$. (d) CTRW at $t=20 000$.
}
\end{figure}

Figure \ref{worldmap}(a) shows the actual hit map
\begin{equation}
h^{\text{r}}({\bf x}\vert t,R)\equiv {h({\bf x}\vert t,R)\over h_0({\bf x}|t,R)}
\end{equation}
for the QTM sample at $t=6 200\simeq 3\tau_{\rm typ}$, where $50\%$ of the launched particles 
have arrived. At this time, maximum density variation on the surface reaches close to 50$\%$.
Figure \ref{worldmap}(b) shows the hit map of the same QTM sample
at a much later time $t=20 000\simeq 10\tau_{\rm typ}$, where about $80\%$ of the particles have arrived.
Although the amplitude of the density variation has decreased from Fig. \ref{worldmap}(a), the pattern of
high and low densities resemble each other.
Figures \ref{worldmap}(c) and (d) show the corresponding maps under the CTRW dynamics,
where the density fluctuations are much weaker.

\section{Summary and conclusions}

In this paper, we have shown that the QTM above its critical dimension $d_c=2$ still differs from its
annealed counterpart CTRW in the first-passage-time statistics. For power-law distributed sojourn times 
at $\mu<1$ where the particle motion becomes subdiffusive, the tail of the FPT distribution for a given sample is 
well-approximated by a sum of exponentials whose time constants are associated with the deepest traps in the system.
Intuitively, this behavior can be understood from rare-event statistics where the passage time of each trajectory
is dominated by the strongest trap visited by the walker. By grouping trajectories that go through the same trap,
one obtains individual terms in the FPT distribution, with weights and a renormalized time constant 
that can be calculated by applying the theory of lattice random walk. The largest time constant in a given
sample of linear size $R$ scales with $R$ as $\tau_{\rm Q}\simeq R^{d\over\mu}$. 
The cutoff time constant $\tau_{\rm typ}\simeq R^{2/\mu}$ for traps that contribute to the
summation is the typical FPT in such a system.

Detailed comparison of our analytic expressions against
simulations of the QTM shows quantitative agreement in the tail of the FPT distribution.
For the peak part of the FPT distribution, however, our treatment is not sufficient to
produce quantitatively accurate results, though the predicted scaling with the system size $R$
is well-obeyed by simulation data. In the future one may consider improvements of the single-trap 
approximation by, e.g., replacing contributions from less strong traps by a annealed average with a
suitable cutoff, as in the solution of the Random Energy Model\cite{Derrida81}.

The QTM differs from the CTRW 
also in the spatially-resolved arrival probability on the absorbing boundary,
whose relative variation can be as big as $50\%$.
It would be interesting to see if such hit maps can be measured in experiments on, say the
exit statistics of mRNAs through nuclear pores. With sufficient statistics, 
the exit pattern may enable detection of large-scale movement of chromosomes inside the cell nucleus.

The work was supported in part by the NSFC under grant U1430237, and by the Research Grants Council of the Hong Kong
SAR under grant 201910.

\appendix

\section{Rare-event-dominated passage time}
\label{sec.fpt}

To justify the replacement of the total elapsed time of a given passage by the largest sojourn time on the path
in our treatment of the CTRW,
let us compare the distribution of the sum $t_{\rm sum}=\sum_{i=1}^N t_i$ and that of the largest term in the sum, 
$t_{\rm max}={\rm max}_i\{t_i\}$. Denoting by $P(t)$ and $f_N(t)$ the distributions of 
$t_i$ and $t_{\rm sum}$, respectively, we have,
\begin{equation}
f_N(t)=\langle\delta\Bigl(t-\sum_{i=1}^Nt_i\Bigr)\rangle,
\end{equation}
where angular brackets denote averaging over the distribution of the $t_i$'s.
Their Laplace transforms are given respectively by,
\begin{eqnarray}
\hat{P}(s)&=&\int_0^\infty P(t)e^{-st}dt=\langle e^{-st}\rangle,\label{hatP}\\
\hat{f}_N(s)&=&\int_0^\infty \langle\delta\Bigl(t-\sum_{i=1}^Nt_i\Bigr)\rangle e^{-st}dt=[\hat{P}(s)]^N.
\label{hat_f}
\end{eqnarray}

Consider now the power-law distribution Eq. (2) at $\omega_0=1$.
For $\mu<1$, its Laplace transform is given by,
\begin{equation}
\hat{P}(s)=1-\Gamma(1-\mu)s^\mu+\mu\sum_{n=1}^\infty {(-1)^{n-1}\over n!}{s^n\over n-\mu}.
\end{equation}
For $Ns^\mu\ll 1$, we have $\hat P(s)\simeq 1-\Gamma(1-\mu)s^\mu$ and $\hat f_N(s)\simeq 1-\Gamma(1-\mu)Ns^\mu$.
Hence we expect $f_N(t)\simeq N\mu t^{-\mu-1}$ for $t\gg N^{1/\mu}$, in agreement with Eq. (\ref{tauextre}).

In the QTM, the sojourn times $t_i$ on the path each satisfies its own distribution
\begin{equation}
P_i(t_i)=\tau_i^{-1}\exp(-t/\tau_i)
\label{P_i}
\end{equation}
with a site-dependent time constant $\tau_i$,  the Laplace transform (\ref{hat_f}) is modified to,
\begin{equation}
\hat{f}_\Gamma(s)=\prod_{i\in\Gamma}{1\over 1+\tau_i s}.
\label{hat_f_gamma}
\end{equation}
For small $s$, we have $\hat{f}_\Gamma(s)\simeq 1-\tau_\Gamma s$
where $\tau_\Gamma=\sum_{i\in\Gamma}\tau_i$. Hence $f_\Gamma(t)$ decays exponentially
at large $t$ with a time constant $\tau_\Gamma$. In the case when $\tau_i$'s are distributed according to 
Eq. (\ref{sft}), $\tau_\Gamma$ can be approximated by the largest time constant $\tau_{\rm max,\Gamma}$
on the path. Multiple visits to the deepest trap can also be treated as in Appendix B.

\section{The single trap model}

In the Main Text we introduced a single trap model where the hopping rate $k_i=\tau_0^{-1}$
except at the trap site located at a distance $r$ from the center of a sphere of radius $R$, where $k_i=\tau_{\rm Q}^{-1}$. 
We shall assume $\tau_{\rm Q}\gg \tau_0 (R/a)^2$ so that the tail of the FPT distribution is dominated by trajectories
that make at least one visit to the trap. Here $a$ is the typical distance travelled by the walker in a time $\tau_0$ when outside the trap.

Let us first revisit the problem of discrete time random walk on a $d$-dimensional hypercubic lattice, paying
special attention to the presence of an absorbing boundary $S$. In each step,
the walker moves from its current site to one of its neighbors with equal probability.
The probability $P({\bf x},n)$ that the walker is at site {\bf x} after $n$ steps satisfies the
lattice diffusion equation,
\begin{equation}
P({\bf x},n+1)={1\over 2d}\sum_{{\bf x}'\in{\rm n.n.\; of\; }{\bf x}}P({\bf x}',n).
\label{lattice-diff}
\end{equation}

Introduce 
\begin{equation}
\Phi({\bf x})\equiv\sum_{n=0}^\infty P({\bf x},n).
\label{Sum_P}
\end{equation}
For a walker launched from the origin, summing both sides of Eq. (\ref{lattice-diff}) over $n$ yields,
\begin{equation}
{1\over 2d}\sum_{{\bf x}'\in{\rm n.n.\; of\; }{\bf x}}\Phi({\bf x}')-\Phi({\bf x})=-\delta_{{\bf x},0}.
\label{lattice-Poisson}
\end{equation}
Here $\delta_{{\bf x},0}=1$ if {\bf x} is at the launch site and zero otherwise.
Equation (\ref{lattice-Poisson}) is a lattice version of the Poisson equation whose continuum limit takes the form,
\begin{equation}
{1\over 2d}\nabla^2\Phi=-\delta({\bf x}).
\label{Poisson}
\end{equation}
The solution of Eq. (\ref{Poisson}) with the boundary $\Phi(|{\bf x}|=R)=0$ is given by,
\begin{equation}
\Phi({\bf x})={2d\over S_d}{1\over d-2}\Bigl({1\over |{\bf x}|^{d-2}}-{1\over R^{d-2}}\Bigr).
\label{Phi-sol}
\end{equation}
Here $S_d=2\pi^{d/2}/\Gamma(d/2)$ is the surface area of a unit sphere in $d$ dimensions.

We now compute the probability $Q({\bf x},n)$ that the walker, launched from the origin, 
arrives at site {\bf x} for the first time in $n$ steps.
Quite generally, we may write,
\begin{equation}
P({\bf x},n)=\sum_{n'=1}^n Q({\bf x},n')U({\bf x},n-n').\qquad (n>0)
\label{lattice-FPT}
\end{equation}
Here $U({\bf x},n)$ is the probability that a walker launched from {\bf x} returns to the same site in $n$ steps.
Summing both sides of Eq. (\ref{lattice-FPT}) over $n$, we obtain,
\begin{equation}
\Phi({\bf x})=\delta_{{\bf x},0}+\Psi({\bf x})\sum_{n=0}^\infty U({\bf x},n),
\label{convolution}
\end{equation}
where
\begin{equation}
\Psi({\bf x})\equiv \sum_{n=1}^\infty Q({\bf x},n)
\label{W-def}
\end{equation}
is the probability that the walker visits site {\bf x} before being absorbed by the boundary $S$.
Applying Eq. (\ref{convolution}) to the site ${\bf x}=0$, we obtain the probability of return,
\begin{equation}
\Psi(0)=1-{1\over\Phi(0)}.
\label{f_d}
\end{equation}

For $d>2$, $P_0\equiv\lim_{R\rightarrow\infty}\Phi(0)$ is finite. Hence the P\'{o}lya's random walk constant
$f_d\equiv\Psi(0)=1-1/P_0$  is less than one. Equivalently, the probability of no return $e_d=1-f_d>0$.
In general, the probability of precisely $k$ returns is given by, 
\begin{equation}
f_{d,k}=(f_d)^ke_d.
\label{k-return}
\end{equation}

When the absorbing boundary $S$ is at a finite distance, the above results acquire a correction which essentially
goes down as $\xi^{2-d}$ where $\xi$ is the distance to the nearest point on the boundary. Ignoring such corrections,
we may approximate $U({\bf x},n)$ by $P(0,n)$.
Consequently, Eq. (\ref{convolution}) yields,
\begin{equation}
\Psi({\bf x})\simeq {\Phi({\bf x})\over\Phi(0)}\equiv w(|{\bf x}|,R).\qquad ({\bf x}\neq 0)
\label{Psi-x}
\end{equation}
Here
\begin{equation}
w(r,R)=(1-f_d){2d\over S_d}{1\over d-2}\Bigl({1\over r^{d-2}}-{1\over R^{d-2}}\Bigr)
\label{w_r}
\end{equation}
is the probability to visit a site at distance $r$ from the launch site, in the presence of the absorbing sphere
of radius $R$. Going back to the lattice walk, the divergence of Eq. (\ref{w_r}) at small distances should be truncated
when $r$ becomes comparable to the lattice constant.

We now return to the single-trap model where the sojourn time $t$ spent by the walker upon each visit to the trap
satisfies the distribution:
\begin{equation}
p(t)=\tau_{\text{Q}}^{-1}\exp(-t/\tau_{\text{Q}}).
\label{single-trap-time}
\end{equation}
Under the assumption that $\tau_{\rm Q}$ is much greater than the typical FPT to the boundary when trap
is not visited, we may write the FPT distribution of the single-trap model as,
\begin{equation}
F_1(t)\simeq [1-w(r,R)]\delta(t)+\sum_{k=1}^\infty w_k(r,R)\Bigl\langle \delta\Bigl(t-\sum_{j=1}^k t_j\Bigr)\Bigr\rangle.
\label{P_1}
\end{equation}
Here the angular bracket indicates average over the sojourn times $\{t_j\}$ distributed according 
to (\ref{single-trap-time}), and
\begin{equation}
w_k(r,R)=(f_d)^{k-1}e_d w(r,R),\qquad (k>0)
\label{gn}
\end{equation}
is the probability that the trap is visited exactly $k$ times before the walker reaches the boundary $S$.

The Laplace transform of Eq. (\ref{P_1}) is given by,
\begin{equation}
\hat{F}_1(s)=1-w(r,R)+\sum_{k=1}^\infty w_k(r,R)\prod_{j=1}^k\langle e^{-st_j}\rangle.
\label{F_1-Laplace}
\end{equation}
From Eq. (\ref{single-trap-time}) we obtain $\langle e^{-st}\rangle =1/(1+s\tau_{\rm Q})$. With the help of Eq. (\ref{gn})
we then have,
\begin{equation}
\hat{F}_1(s)=1-w(r,R)+e_dw(r,R){1\over e_d+s\tau_{\rm Q}}.
\label{F_1-Laplace_a}
\end{equation}
Consequently,
\begin{equation}
F_1(t)\simeq [1-w(r,R)]\delta(t)+w(r,R){e_d\over\tau_Q}\exp(-e_dt/\tau_{\rm Q}).
\end{equation}
See also Ref. \cite{benichou10} for a general discussion on the decomposition. 
Integrating the tail of the distribution, we obtain Eq. (\ref{cqexp}).

\section{Statistics of $C(t,R)$ in the QTM}

The sum in Eq. (\ref{C_superposition}) is restricted to sites with $\tau_i>\tau_{\rm typ}(R)$. 
Alternatively, we may extend the sum to all sites
inside the sphere, where a given site $i$ contributes a term
\begin{equation}
c_i(t)=\begin{cases}
w(r_i,R)e^{-e_d t/\tau_i}, & \text{if $\tau_i>\tau_{\rm typ}(R)$;}\\
0,& \text{otherwise.}
\end{cases}
\label{c-i}
\end{equation}
The mean and variance of $c_i$ are given by,
\begin{eqnarray}
\langle c_i(t)\rangle&=&\int_{\tau_{\rm typ}(R)}^\infty d\tau {\mu\over \tau^{\mu+1}}w(r_i,R)e^{-e_d t/\tau}\nonumber\\
&=&t^{-\mu}w(r_i,R){\mu\over e_d^{\mu}}\gamma\Bigl(\mu,{e_dt\over \tau_{\rm typ}(R)}\Bigr),
\end{eqnarray}
\begin{eqnarray}
\langle [\Delta c_i(t)]^2\rangle&=&w^2(r_i,R)\Bigl[ \int_{\tau_{\rm typ}(R)}^\infty d\tau {\mu\over\tau^{\mu+1}}
e^{-2e_d t/\tau}\nonumber\\
&&-\Bigl(\int_{\tau_{\rm typ}(R)}^\infty d\tau {\mu\over \tau^{\mu+1}}e^{-e_d t/\tau}\Bigr)^2\Bigr]\nonumber\\
&=&t^{-\mu}w^2(r_i,R)\Bigl[ {\mu\over (2e_d)^{\mu}}\gamma\Bigl(\mu,{2e_dt\over \tau_{\rm typ}(R)}\Bigr)\nonumber\\
&&-t^{-\mu}{\mu^2\over e_d^{2\mu}}\gamma^2\Bigl(\mu,{e_dt\over \tau_{\rm typ}(R)}\Bigr)\Bigr].
\end{eqnarray}
Here $\gamma(\mu,z)=\int_0^zdx x^{\mu-1}e^{-x}$ is the incomplete gamma function.
Since the $\tau_i$'s are independently chosen from the distribution Eq. (2), the $c_i$'s are also statistically independent.
Hence the mean and variance of $C(t,R)=\sum_i c_i(t)$ are given by,
\begin{equation}
\langle C(t,R)\rangle\simeq\int_{|{\bf x}|<R} 
d^d{\bf x} t^{-\mu}w(|{\bf x}|,R){\mu\over e_d^{\mu}}\gamma\Bigl(\mu,{e_dt\over \tau_{\rm typ}(R)}\Bigr).
\label{mean_C}
\end{equation}
\begin{eqnarray}
\langle [\Delta C(t,R)]^2
&\simeq&\int_{|{\bf x}|<R} d^d{\bf x} t^{-\mu}w^2(|{\bf x}|,R)\nonumber\\
&&\times{\mu\over (2e_d)^{\mu}}\gamma\Bigl(\mu,{2e_dt\over \tau_{\rm typ}(R)}\Bigr),
\label{Delta_C}
\end{eqnarray}
where we have replaced summation over $i$ by integral over space inside the sphere.
For $2<d<4$, carrying out the integrals over {\bf x} yield Eqs. (\ref{C_mean}) and (\ref{C_var}).
For $d>4$, the integral over ${\bf x}$ in (\ref{Delta_C}) is dominated by contributions close to the origin,
and hence the lattice cut-off should be considered.

\end{document}